\documentclass{article}
            \renewcommand{\a}{\alpha}
            
            \newcommand{\g}{\gamma}
           \usepackage{amsfonts}
            \usepackage{graphics}
            \begin{document}
           \title{A Possible Approach to Inclusion of Space and Time in Frame Fields of
           Quantum Representations of Real and Complex Numbers}
          \author{Paul Benioff \\
            Physics Division, Argonne National
           Laboratory,\\
           Argonne, IL 60439, USA \\
           email: pbenioff@anl.gov}
           \maketitle

            \begin{abstract}
           This work is based on the field of reference frames
           based on quantum representations of real and complex
           numbers described in other work. Here frame domains
           are expanded to include space and time lattices.
           Strings of qukits are described as hybrid systems
           as they are both mathematical and physical systems.
           As mathematical systems they represent numbers.
           As physical systems in each frame the strings have
           a discrete Schrodinger dynamics on the lattices. The
           frame field has an iterative structure such that the
           contents of a stage j frame have images in a stage j-1
           (parent) frame. A discussion of parent frame images
           includes the proposal that points of stage j frame
           lattices have images as hybrid systems in parent frames.
           The resulting association of energy with images of lattice
           point locations, as hybrid systems states, is discussed.
           Representations and images of other physical systems in
           the different frames are also described.
           \end{abstract}

           \section{Introduction}
           The need for a coherent theory of physics and mathematics
           together arises from considerations of the basic relationship
           between physics and mathematics. Why is mathematics relevant
           to physics \cite{Wigner,Hamming,Boniolo}. One way to
           see the problem is based on the widely held Platonic
           view of mathematics. If mathematical systems have an
           abstract ideal existence outside of space and time and
           physics describes the property of systems in space and
           time, then why should the two be related at all?  Yet
           it is clear that they are very closely related.

           The problem of the relationship between the foundations
           of mathematics and physics is not new.  Some recent work
           on the subject is described in\cite{Davies,Bendaniel,Wolfram}
           and in \cite{Tegmark1,Tegmark,Bernal}. In particular the  work of
           Tegmark \cite{Tegmark} is is quite explicit in that it
           suggests that the physical universe is a mathematical  universe.

           Another approach to this problem is to work towards
           construction of a theory that treats physics
           and mathematics together as one coherent whole
           \cite{BenTCTPM,BenTCTPMTEC}. Such a theory would be
           expected to show why mathematics is so important to
           physics by describing details of the relation between
           mathematical and physical systems.

           In  this paper a possible approach to a coherent theory of
           physics and mathematics is described.  The
           approach is based on the field of reference frames that
           follows from the properties of quantum mechanical
           representations of real and complex numbers
           \cite{BenRRCNQT,BenRFFQTRRCN}.

           The use, here, of reference frames is similar in many ways
           to that used by different workers in areas of physics
           \cite{Aharonov,Bartlett,vanEnk,Enk,Poulin,Bartlett1}. In
           general, a reference frame provides a background or basis
           for descriptions of systems. In special relativity,
         reference frames for describing physical dynamics are
         based on choices of inertial coordinate systems. In quantum
         cryptography,  polarization directions are used
         to define reference frames for sending and receiving messages
         encoded in  qubit string states.

         The use of reference frames here  differs from those
         noted above in that the frames are not based on a preexisting space
         and time as a background. Instead they  are
         based on a mathematical parameterization of quantum theory
         representations of real and
         complex  numbers. In particular, each frame $F_{j,k,g}$ in the
         field is based on a quantum theory representation,
         $R_{j,k,g},C_{j,k,g},$ of the real and complex numbers
         where $R_{j,k,g}$ can be viewed as a set of equivalence classes of
         Cauchy sequences of  quantum states of qukit strings.
         $C_{j,k,g}$ is a set of pairs of these equivalence classes.
         The parameter $k\geq 2$ is the base ($k=2$ for qubits), $g$
         denotes a basis choice for the states of qukit strings that are
         values of rational numbers, and $j$ denotes an iteration stage.
         The existence of iterations follows from the observation that
         the representations of real and complex numbers are based on
         qukit string states. These are elements of a Hilbert space that
         is itself defined as a vector space over a field of complex
         numbers. Consequently one can use the real and complex
         numbers constructed in a stage $j$ frame as the base of a
         stage $j+1$ frame.

         Each reference frame contains a considerable number of mathematical
         structures. Besides $R_{j,k,g}$ and $C_{j,k,g},$  a frame
         $F_{j,k,g}$ contains representations of all mathematical
         systems that can be described as structures based on
         $R_{j,k,g}$ and $C_{j,k,g}.$  However frames do not contain
         physical theories as mathematical structures based on
         $R_{j,k,g},C_{j,k,g}.$ The reason is that the frames do not
         contain any representations of space and time.

         The goal of this paper is to take a first step in remedying
         this defect by expansion of the domain of each frame
         to include discrete space and time lattices. The lattices,
         $L_{j,k,L,m},$ in a frame, $F_{j,k,g},$ are such
         that the number of points in each dimension is given by
         $k^{L},$ the spacing $\Delta=k^{-m}$ and $m=0,1,\cdots,L.$
         For each lattice, $L$ and $m$ are fixed with $L$ an arbitrary
         nonnegative integer. It follows that each dimension
         component of the  location of each point in a lattice is a
         rational number expressible as a finite string of base $k$
         digits.

         Representations of physical systems of different types
         are also present in each frame. However, the emphasis
         here is on strings of qukits, $S_{j,k',L',m'},$ present in
         each frame. These strings are considered to be hybrid
         systems in that they are both physical systems and
         mathematical systems. As mathematical systems,
         the quantum states of each string, in some basis,
         represent a set of rational numbers. As physical systems
         the motion of strings in a frame is described relative to a
         space and time lattice in the frame. This dual role is
         somewhat similar to the concept that information is
         physical \cite{Landauer}.

         Considerable space in the paper is devoted to how observers
         in a stage $j-1$ (parent) frame view the contents of a
         stage $j$ frame. For an observer, $O_{j},$ in a frame
         $F_{j,k,g}$ the numbers in the real and complex number
         base of the frame are abstract and have no structure.
         The only requirement is that they satisfy the set of
         axioms for real or complex number systems. Points of
         lattices $L_{j,k,L,m}$ in the frame are also regarded
         as abstract and without structure. The only requirement
         is that the lattices satisfy some relevant geometric axioms.

         The view of the contents of a stage $j$ frame as seen by an observer
         $,O_{j-1},$ in a stage $j-1$ frame, $F_{j-1,k',g'},$ is quite
         different. Elements of the stage $j$ frame that $O_{j}$
         sees as abstract and with no structure, are seen by $O_{j-1}$
         to have structure. Numbers in $R_{j,k,g}$ are seen by
         $O_{j-1}$ to be equivalence classes of Cauchy sequences of
         states of stage $j-1$ hybrid systems. Space points of stage
         $j$ lattices with $D$ space and one time dimension  are
         seen in a stage j-1 frame to be $D$ tuples of hybrid systems
         with the location of each point given by a state of the
         $D$ tuple. Time points are seen to be hybrid systems whose
         states correspond to the possible lattice time values.

         All this and more is discussed in the rest of the paper.
         The next section is a brief summary of quantum theory
         representations and the resulting frame fields
         \cite{BenRRCNQT,BenRFFQTRRCN}. Section \ref{PFCTPM}
         describes a possible approach to a coherent theory of
         physics and mathematics as the inclusion of space and
         time lattices in each frame of the frame field. Properties
         of the lattices in the frames are described. Qukit strings
         as hybrid systems are discussed in the next section. Their
         mathematical properties as rational number systems with
         states as values of rational numbers are described. Also
         included  is a  general Hamiltonian description of the
         rational number states as energy eigenvalues and a
         Schr\"{o}dinger equation description of the dynamics of
         these systems.

         Section \ref{FEVPF} describes frame entities as viewed from
         a parent frame. Included is a description of real and
         complex numbers, quantum states and Hilbert spaces, and
         space and time lattices.  Section \ref{LHS} discusses in
         more detail  a stage $j-1$ views of stage $j$ lattice points
         and locations as tuples of hybrid systems and point
         locations as states of the tuples. Dynamics of these tuples
         in stage $j-1$  is briefly described as is the parent
         frame view of the dynamics of physical systems in general.

         The last section is a discussion of several points.  The
         most important one is that frame field description given
         here leads to a field of different descriptions of the
         physical universe, one for each frame,  whereas there
         is just one. This leads to the need to find some way to
         merge or collapse the frame field to correspond to the
         accepted view of the physical universe. This is discussed
         in the section as are some other points.

           Whatever one thinks of the ideas and systems described in
           this work, it is good to keep the following points in
           mind. One point is that the existence of the reference frame
           field  is based on properties of states of qukit string
           systems representing values of rational numbers. However
           the presence of a frame field is more general in that it
           is not limited to states of qukit strings. Reference frame
           fields arise for any quantum representation of rational
           numbers where the values of the rational numbers, as
           states of some system, are elements of a vector space over
           the field of complex numbers.

           Another point is that the three dimensional reference
           frame field described here exists only for quantum
           theory representations of the natural numbers, the integers,
           and the rational numbers. Neither the
           basis degree of freedom $g$, nor the iteration stages, $j,$
           are present in classical representations. This is the case even
           for classical representations based on  base $k$ digit or kit
           strings.  The reason is that states of digit strings are not
           elements of a vector space over a complex number field.

           Finally, although understandable, it is somewhat of a
           mystery why so much effort in physics has gone into the
           description of various aspects of quantum geometry and
           space time and so little into quantum representations of
           numbers.  This is especially the case when one considers
           that natural numbers, integers, rational numbers, and
           probably real and complex numbers, are even more fundamental
           to physics than is geometry.

         \section{Review: Quantum Theory Representations of
         Numbers and Frame Fields}\label{R}
         \subsection{Quantum Theory Representations of
         Numbers}\label{QTRN}
         \subsubsection{Representations of Natural Numbers,
         Integers, and Rational Numbers}\label{RNNIRN}
          In earlier work \cite{BenRRCNQT,BenRFFQTRRCN} quantum
          theory representations of natural numbers $(N)$, integers $(I)$,
          and rational numbers $(Ra)$,  were described by  states
          of finite length qukit strings that include one qubit.
          To keep the description as simple as possible, the strings
          are considered to be finite sets of qukits and one qubit
          with the qukits and qubit parameterized by integer labels.
          The natural ordering of the integers serves to order the set
          into a string.

          This purely mathematical representation of qukit strings
          makes no use of physical representation of qukit strings
          as extended systems in space and/or time. Physical
          representations are described later on in Section
          \ref{QSSHS} after the introduction of space and time
          lattices into each frame of the frame field.

          The qukit ($q_{k}$) string states are given by
          $|\g,s\rangle_{k,L,m,g}$ where $s$ is a $0,1,\cdots,k-1$
          valued function with domain $0,\cdots,L-1,$ and
           $\g = +,-$ denotes the sign. The string location of the sign
           qubit is given by $m$ where $m=0,\cdots, L.$ $L$ is any nonnegative
           integer. This expresses the range of possible locations of the
           sign qubit from one end of the string to the other. By
           convention $m=0$ has the sign qubit at the right end
           of the string, $m=L$ at the left end. The
           qubit can occupy the same integer location as a qukit.
           The reason for the subscript $g$ will be clarified later on.

           A compact notation is used where the location  $m$ of the
           sign qubit is also the location of the $k-al$ point. As
           examples, the base $10$ numbers $3720,$ $-.0474,$ $ -12.71$
           are represented here by $|3720+\rangle,$ $|-0474\rangle,$
           $|12-71\rangle$ respectively.

            Strings are characterized by the values of $k,L,m.$ For each
            $k,$ $L,$ and $m$ the string states $|\g,s\rangle_{k,L,m,g}$
            give a unified quantum theory representation of natural numbers
            and integers in
            $N_{k,L},$ $I_{k,L},$ and $Ra_{k,L,m}.$ For numbers in $N_{k.L},$ $\g=+, m=0$;
            for numbers in $I_{k,L}$, $m=0,$  and there are no restrictions
            for $Ra_{k,L,m}.$ Here $Ra_{k,L,m}$ is the set of rational
            numbers expressible as $l\Delta$ where $l$ is any integer
            whose absolute value is $<k^{L}$ and $\Delta =k^{-m}.$ $I_{k,L}=
            Ra_{k,L,0}$ and $N_{k,L}$ is the set of nonnegative integers
            in $I_{k,L}.$ The correspondence between the numbers $l\Delta$
            and the states $|\g,s\rangle_{k,L,m,g}$ is given by the
            observation that each $s$ corresponds to an integer
            $l=\sum_{j=0}^{L-1}s(j)k^{j}.$ Also, as noted, $m$ is the
            location of the $k-al$ point measured from the right end
            of the string.

           Since one is dealing with quantum states of qukit
           strings,  states with leading and trailing $0s$ are
           included. In this case there are many states that
           are all arithmetically equal even though they are
           orthogonal quantum mechanically. For example
           $|013-470\rangle =_{A}|13-47\rangle$ even though the two
           states are orthogonal.

           The set of states so defined form a basis set that spans
           a Fock  space $\mathcal{F}_{k}$ of states. A Fock space is
           used because the basis set includes states of $q_{k}$
           strings of different lengths.\footnote{Representations of
           these states by use of qukit annihilation creation operators
           will not be done here as it is not needed for this paper.
           Representations in terms of these operators for different
           types of numbers are described in \cite{BenRNQM} and
           \cite{BenRRCNQT}.  Also see \cite{Raffa}.}  Linear
           superposition states  in
           the space have the form\begin{equation}\label{linsuppsi}
           \psi=\sum_{L,m}\sum_{\g,s}c_{\g,s,L,m}|\g,s\rangle_{k,L,m,g}
           \end{equation} The $L$ and $m$ sums are over all positive
           integers and  from $0$ to $L$, and the $s$ sum is over
           all functions with domain $[0,L-1].$

           As already noted, the states $|\g,s\rangle_{k,L,m,g}$ of
           $q_{k}$ strings are values of rational numbers.
           Quantum mechanically they also represent a basis choice
           of states in the Fock space $\mathcal{F}_{k}.$ However
           the choice of a basis is arbitrary in that there are an
           infinite number of possible choices.  Here the choice of
           a basis is parameterized by the variable $g.$ Since
           choice of a basis is equivalent to fixing a gauge, $g$ is
           also a gauge fixing parameter.

           One can also describe gauge and base transformation
           operators on these states. Gauge transformations correspond
           to a basis change ($g$ to $g'$) and base
           transformations correspond to a base change ($k$ to
           $k'$). Details are given in \cite{BenRFFQTRRCN}.

         \subsubsection{Real and Complex Numbers}
         Quantum theory representations of real numbers are defined
         here as equivalence classes of Cauchy sequences of states
         of finite $q_{k}$ strings that are values of rational
         numbers. Let $\psi$ be a function on the natural numbers
         such that for each $n$ $\psi(n)$ is a basis state:
         \begin{equation}\label{psinunln}\psi(n)=|\g_{n},s_{n}
         \rangle_{k,L_{n},m_{n},g}. \end{equation} For each $n$
         the interval $[0,L_{n}-1]$ is the domain of $s_{n}.$

         The sequence $\psi$ is a Cauchy sequence if it satisfies the
         Cauchy condition:\begin{equation}\label{cauchy}\begin{array}{l}
         \hspace{0.5cm}\mbox{ For each $\ell$ there is a $p$
         where for all $j,h>p$} \\ \hspace{1cm}
         |(|\psi(j)-_{A,k,g}\psi(h)|_{A,k,g})\rangle_{k,g}
         <_{A,k,g}|+,-\ell\rangle_{k,g}.\end{array} \end{equation}

          Here $|(|\psi(j)-_{A,k,g}\psi(h)|_{A,k,g})
         \rangle_{k,g}$ is the basis state that is the base $k$ arithmetic
         absolute value of the state resulting from the arithmetic
         subtraction of $\psi(h)$ from $\psi(j).$ The Cauchy condition
         says that this state is arithmetically less than or equal to the
         state $|+,-\ell\rangle_{k,g}=|+,0_{[0,-\ell+1]}1_{-\ell}\rangle_{k,g}$
         for all $j,h$ greater than some $p$. Here $|+,-\ell\rangle$
         is a string state that represents the number $k^{-\ell}.$
         The subscripts $A,k,g$ in the definition of the Cauchy
         condition indicate that the operations are arithmetic and
         are defined for base $k$ string states in the gauge $g$.
         They are \emph{not} the usual quantum theory operations.

         The definition can be extended to sequences $\psi$ of
         linear superpositions of basis states.  In this case one
         defines a probability $P_{j,h,\ell}$  as a sum over all
         components of $\psi(j)$ and $\psi(h)$ that satisfy the
         second line of Eq. \ref{cauchy}. The state $\psi$ is Cauchy if the
         probability $P_{\psi}=1$ where\begin{equation}\label{limPpsi}
         P_{\psi}=\liminf_{\ell\rightarrow\infty}\limsup_{p\rightarrow
         \infty}\inf_{j,h>p}P_{j,h,\ell}.\end{equation}

         Two sequences $\psi$ and $\psi'$ are equivalent, $\psi\equiv\psi',$
         if the sequence defined by the termwise arithmetic
         difference of $\psi$ and $\psi'$ converges to $0$. The
         specific condition for this is expressed by Eq.
         \ref{cauchy} if one replaces $\psi(h)$ with $\psi'(h).$ The
         relation $\psi\equiv\psi'$ is used to define equivalence
         classes of Cauchy sequences of $q_{k}$ string states.
         The set of these equivalence classes is a quantum theory
         representation $R_{k,g}$ of the real numbers.

         Quantum equivalence classes of Cauchy sequences of states  are
         larger than classical equivalence classes because each
         quantum equivalence class contains many sequences of states
         that have no classical correspondent. However no new
         equivalence classes appear. This is a consequence of the
         fact that each quantum equivalence class contains a
         basis valued sequence that corresponds to a classical
         sequence of finite base $k$ digit strings.

         One can also define a canonical representation of each
         equivalence class as a sequence $\psi(n)$ of basis states
         such that if $m>n$ then $\psi(n)$ is an initial part of
         $\psi(m).$  This is similar to the usual classical
         canonical representation of an equivalence class of real
         numbers as an infinite string of digits with a $k-al$
         point. The quantum representation would be an infinite
         string of qukits with each qukit in one of $k$ basis
         states.

         Extension of the above to include quantum representations
         of complex numbers is straightforward. One method represents
         complex rational numbers by pairs of states
         of finite $q_{k}$ strings. This what is actually done in
         computations involving complex numbers. The state components
         of the pair represent values of real and imaginary rational numbers.

         Cauchy sequences of these state pairs are defined by applying
         the Cauchy condition separately to the component sequences
         of real rational number states and imaginary rational number
         states.  Two Cauchy sequences $\psi$ and $\psi'$ of state
         pairs are equivalent, $\psi\equiv\psi',$ if the termwise
         arithmetic differences  of the real parts, $Re\psi-_{A}Re\psi',$
         and of the imaginary parts, $Im\psi-_{A}Im\psi'$  each converge
         to $0.$ The resulting set of equivalence classes of Cauchy
         sequences is a quantum theory representation of
         the set of complex numbers.

         \subsection{Fields of Iterated Reference
         Frames}\label{FIRF}
         Quantum theory representations of real and complex numbers
         differ from classical representations in two important
         ways.  One is the presence of the gauge freedom or basis
         choice freedom. This is indicated by the $g$ subscript in
         $R_{k,g},C_{k,g}.$

         The other difference is based on the observation that states of
         qukit strings are elements of a Hilbert space or a Fock
         space. From a mathematical point of view these spaces are
         vector spaces over the fields of real and complex numbers.
         It follows from what has been shown that $q_{k}$ string
         states, as elements of a vector space over the field of real
         and complex numbers, can be used  to construct
         other representations of real and complex numbers. This
         suggests the possibility of iteration of the construction
         described here as the quantum representations of real and
         complex numbers can in turn be used as the base of Hilbert
         and Fock spaces for a repeated construction. Here the
         iteration stage is another degree of freedom for the frame
         field.

         The third degree of freedom arises from the free choice of
         the base choice for the humber representation.  This choice,
         denoted by the $k$ subscript, is common to both quantum and
         classical representations.

         These three degrees of freedom can be combined to describe a
         three dimensional reference frame field. Each reference frame
         $F_{j,k,g}$ is based on a quantum representation
         $R_{j,k,g},C_{j,k,g}$ of the real and complex numbers. The
         subscripts denote the iteration stage $j$, the base $k$,
         and the gauge $g$. Each reference frame contains
         representations of Hilbert and Fock spaces as mathematical
         structures over $R_{j,k,g},C_{j,k,g}.$

         Because the iteration degree of freedom is directed, it is
         useful to use genealogical terms to describe the iteration stages.
         Frames that are ancestors to a given frame $F_{j,k,g}$ occur
         at stages $j'$ where $j'<j.$ Frames that are descendants
         occur at stages $j'$ where $j'>j.$

         From a mathematical
         point of view there are several possibilities for the
         stages:  The frame fields can have a finite number of
         stages with both a common ancestor frame and a set of
         terminal frames. The fields can also be one way infinite
         with either a common ancestor and no terminal stage, or
         with a terminal stage and no common ancestor; or they can
         be two way infinite.  They can also be  cyclic. These last
         two cases have no common ancestor or terminal frames.

\begin{figure}[ht]
         \begin{center} \resizebox{150pt}{150pt}{\includegraphics
         [280pt,160pt][620pt,520pt]{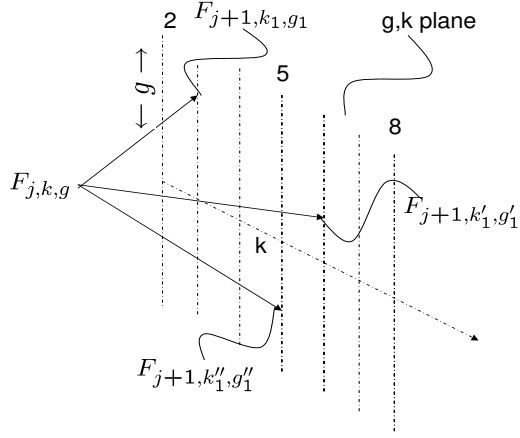}}\end{center}
         \caption{Schematic illustration of frames coming from frame
         $F_{j,k,g}.$ Each of the stage $j+1$ frames is based on
         quantum representations of real and complex numbers as
         equivalence classis of Cauchy sequences of qukit string
         states in $F_{j,k,g}.$ The distinct vertical lines
         in the $k,g$ plane denote the discreteness of the integral
         values of $k\geq 2.$ Only three of the infinitely many
         frames coming from $F_{j,k,g}$ are shown. Here $k$ denotes
         the qukit base and $g$ denotes a gauge or basis choice.}
         \label{PACTPM2} \end{figure}

         Another way to illustrate the frame field structure is to
         show, schematically, frames emanating from frames.
         Figure \ref{PACTPM2} illustrates a slice of the frame field
         for a fixed value of $j.$  Each point $k',g'$ in the $k-g$
         plane denotes a reference frame $F_{j+1,k',g'}$ at
         the next iteration stage with $R_{j+1,k',g'},C_{j+1,k',g'}$
         as its real and complex number base. 

         Three different viewpoints of the real and complex numbers
         as frame bases are of use here. $R_{j,k,g},C_{j,k,g}$ is a
         view from outside the frame field that denotes the position
         of the numbers with respect to the field degrees of
         freedom.  To an observer inside $F_{j,k,g}$ the elements of
         $R_{j,k,g},C_{j,k,g}$ are seen as external,
         abstract, featureless elements. They have no structure
         other than that which follows from the requirement that
         they satisfy the axioms for real and complex
         numbers.  This assumption is made because it is the  view
         held, at least implicitly by physicists.  It also
         corresponds to how numbers are treated in physical
         theories. The only properties of numbers relevant to
         theories are those derived from the appropriate axioms.

         An observer in a parent frame $F_{j-1,k',g'}$ sees the
         elements of $R_{j,k,g},C_{j,k,g}$  as having structure.
         They are seen as equivalence
         classes of Cauchy sequences of states of finite $q_{k}$
         strings.  This is in addition to their having properties
         derived from the relevant axiom sets.

         \section{A Possible Approach to a Coherent Theory of
         Physics and Mathematics}\label{PFCTPM}
         The main consideration of this paper is the proposed use of
         the reference frame field as a possible approach to a
         coherent theory of physics and mathematics. This ensures
         that quantum theory representations of natural numbers,
         integers, rational numbers, and real and complex numbers
         will play a basic role in the theory.

         So far the reference frames contain mathematical systems.
         These include quantum theory representations of numbers,
         qukit strings, and representations of other mathematical
         systems as structures based on the different types of numbers.
         Physical theories and systems  are not present in the
         frames. The reason is that there are no representatios of space
         and time in the frames. These are needed for theories to describe the
         kinematics and dynamics of systems moving in space and time.

         This must be remedied if the frame field is to be an approach
         to a coherent theory. One way to fix this is to
         expand the domains of the frames to include physical
         systems and descriptions of their dynamics.

         \subsection{Space and Time Lattices in Reference
         Frames}\label{STLRF}

         A first step in this direction is to expand the domain of
         each frame in the field to include discrete lattices of
         space and time. The reason for working with discrete
         instead of  continuum space and time will be noted later.

         To be more specific, each frame $F_{j,k,g}$ includes a set
         $\mathcal{L}_{j,k}$ of space and time lattices. Each
         lattice $L_{j,k,M,\Delta}$ in the set is such
         that the number $M$ of points in each dimension is finite
         and the spacing $\Delta$ of points is also finite.
         In this paper the number $D$ of space dimensions in the
         lattice will be arbitrary. To keep things simple, the
         number $M$ of points and spacing $\Delta$ in each of the
         $D$ space dimensions and the time dimension will be
         assumed to be the same.

          It should be noted that, to an observer in a frame, the
          points in each space and time lattice in
          a frame \emph{are} (emphasis on "are") points of space
          and time relative to that frame. They are not merely
          mathematical representations or descriptions of some
          external space and time. In addition an observer in a
          frame sees the points of space and time lattices in in the
          frame as abstract, featureless points with no structure.
          The only requirement is that the lattices should satisfy
          appropriate geometrical axioms.

         A restriction on the lattices in frames is that
         the values of $M$ and $\Delta$ for each space and time
         lattice in the set $\mathcal{L}_{j,k}$ are given by
         \begin{equation}\label{Mdelta}\begin{array}{c}M=k^{L}\\
         \Delta=k^{-m}.\end{array}\end{equation} Here $L$ is an
         arbitrary nonnegative integer, $m=0,1,\cdots,L,$ and $k$ is
         the same integer $\geq 2$ that is the $k$ subscript in the
         frame label, as in $F_{j,k,g}.$ Because of this restriction
         the lattices $L_{j,k,M,\Delta}$ will be labeled from now on
         as $L_{j,k,L,m}.$ Based on Eq. \ref{Mdelta}, the location
         of each space or time point in each dimension  $z$ is given
         by a rational number value, $x_{z}=l_{z}\Delta,$ where
         $l_{z}$ is a nonnegative integer $<M.$

         Even though these requirements might seem  restrictive,
          they are sufficiently general to allow lattices of
          arbitrarily small spacing and arbitrarily many points.
          Also they can be used to describe sequences of lattices
          that become continuous in the limit. An example of such a
          sequence is given by setting $m=[L/2]$ and increasing $L$
          without bound.

          It follows from this description that the points of a lattice,
          $L_{j,k,L,m},$ with $D$ space and one time dimension
          can be taken to be $D+1$-tuples of rational numbers. The
          value of each number in the tuple is  expressible as a finite
          string of base $k$ digits.\footnote{Here, unlike the usual case, the
          frames do not include a continuum space and time as a common
          background for the lattices in $\mathcal{L}_{j,k}.$ One may hope
          that the structure of space and time, as a continuum or as some
          other structure, will emerge when one finds a way to
          merge the frames in the frame field.}

          So far each frame contains in its domain, space and time lattices,
          and strings of qukits  that are numbers. It is reasonable
          to expect that it also contains various types of physical
          systems. For the purposes of this paper the types of included
          physical systems do not play an important role as the main
          emphasis here is on qukit string systems. Also in this first paper
          descriptions of system dynamics will be limited to nonrelativistic
          dynamics.

          It follows from this that each frame includes a
          description of the dynamics of physical systems based on
          the space and time lattices in the frame.  The kinematics and
          dynamics of the systems are expressed by theories that are
          present in the frame as mathematical structures over the
          real and complex number base of the frame.  This is the
          case irrespective of whether the physical systems are
          particles, fields, or strings, or have any other form.

          One reason space and time are described as discrete lattices
          instead of as continua is that it is not clear what the
          appropriate limit of the discrete description is. As
          is well known, there are many different descriptions of
          space and time that are present in the literature. The
          majority of these descriptions arise from the need to
          combine quantum mechanics with general relativity.
          They include use of various quantum geometries
          \cite{QSST,Ashtekar1,Madore,Connes,Horowitz}, and space time as
          a foam \cite{Ng,Ng1,Amelino,Kirillov} and as a spin
          network as in loop quantum gravity \cite{Ashtekar}.
          These are in addition to the often used assumption of a
          fixed flat space and time continuum that serves as a
          background arena for the dynamics of all physical systems,
          from cosmological to microscopic. Space and time may also
          be emergent in an asymptotic sense \cite{EST}.

          The fact that there are many different lattices in a frame
          each characterized by different values of $L$ and $m$
          and that each can serve as a background space and time is
          not a problem.  This is no different than the fact that
          one can use many different space and time lattices with
          different spacings and numbers of points to describe
          discrete dynamics of systems.

         \section{Qukit String Systems as Hybrid Systems}\label{QSSHS}
         So far the domain of each frame contains space and time
         lattices, many types of physical systems (such as
         electrons, nuclei, atoms, etc.) and physical theories as
         mathematical structures based on the real and complex
         numbers. These describe the kinematics and dynamics of
         these systems on the lattices. Also included are qukit strings.
         States of these strings were seen to be values of rational
         numbers.  These were used to describe real and complex
         numbers as Cauchy sequences of these states.

         Here it is proposed to consider these strings as systems
         that can either be numbers, i.e.  mathematical systems,
         or they can be physical systems.  Because of this dual
         role, they are referred to as hybrid systems. As such
         they will be seen to play an important role.

          Support for this proposal is based on the observation
         that the description of qukit strings
         as both numbers and physical systems is not much
         different than the usual view in physics regarding qubits
         and strings of qubits. As a unit of quantum information, the
         states of a qubit can be $|0\rangle$ and $|1\rangle$ which
         denote single digit binary numbers.  The states can also be
         $|\uparrow\rangle, |\downarrow\rangle$ as  spin projection
         states of a physical spin $1/2$ system. In the same way
         strings of qubits are binary numbers in quantum computation,
         or they can represent physical systems such as spins or atoms
         in a linear ion trap \cite{iontrap}. To be blunt about it,
         "Information is Physical" \cite{Landauer}, and information is
         mathematical.

         Also it is reasonable to expect that the domain of a
         coherent theory of physics and mathematics together would
         contain systems that are both mathematical systems and
         physical systems. The hybrid systems are an example
         of this in that they are number systems, which are
         mathematical systems,  and  they are physical systems.

         Let $S_{j,k',L,m}$ denote a hybrid system in $F_{j,k,g}$
         that contains $L$ qukits and one qubit. $L$ is any nonnegative
         integer, $m$ is any nonnegative integer $\leq L,$ $j$ is
         the iteration stage of any frame containing these systems,
         and $k'$ is the base (or dimension of the Hilbert state
         space) of the $q_{k'}$ in the string system. Note that
         $k'$ can be different from the base $k$ of the frame
         $F_{j,k,g}$  containing these systems.

         The  different $q_{k'}$ in the system are distinguished by
         labels in the integer interval $[1,L].$ The qubit has
         label $m$ where $0\leq m\leq L.$ The canonical ordering of
         the integers serves to order the $q_{k'}$ and qubit
         into a string system.

         The presence of the sign qubit is needed if the states of the hybrid
         systems are to be values of rational numbers.  Since the qubit also
         corresponds to the $k'-al$ point, the value of $m$
         gives the location of the sign and $k'-al$ point in $S_{j,k',L,m}.$

         Note that there is a change of emphasis from the usual
         description of numbers.  In the usual description, strings
         of base $k'$ digits, such as $1423.45$ with $k'=10$ are called
         rational numbers. Here states, such as $|1423+45\rangle_{k,g},$
         are  called \emph{values} of rational numbers.  The hybrid system
         will also be referred to as  a rational number system. The reason
         is that the set of all basis states of a hybrid system correspond
         to a set of $k^{L}$ values of rational numbers. The type of number
         $N,I,Ra$, represented  is characterized by the value of
         $m$ and state of the sign qubit in $S_{j,k',L,m}.$\footnote{One
         would like to call $S_{j,k',L,m}$ a
         rational number instead of a rational number system. This
         would agree with the usual physical description of
         systems. For example a physical system of a certain type is
         a proton, not a proton system. However referring to
         $S_{j,k',L,m}$ as a rational number instead of a rational
         number system seems so at odds with the usual use of the
         term that it is not done here.}

         As was noted, the states of the $S_{j,k',L,m}$
         systems in a frame $F_{j,k,g}$  are elements of a
         Hilbert space $\mathcal{H}_{k',L,m}$ in the frame.  The
         choice of a basis set or gauge $g'$ fixes the states of
         $S_{j,k',L,m}$ that are values of rational numbers.
         These states are represented as $|\g,s\rangle_{j,k',L,m,g'}.$
         Often the $L,m,g'$  subscripts on the states will be dropped
         as they will not be needed for the discussion.

         The description of the hybrid systems as strings of qukits
         is one of several possible structures.  For example, as physical
         systems that move and interact on a space lattice $L_{j,k,L'.m'},$
         the strings could be open with free ends or closed loops.
         In this case, aspects of string theory \cite{Zweibach} may be
         useful in describing the physics of the strings.

         Whatever structure the hybrid systems have, it would be
         expected that, as bound systems, they have a spectrum of
         energy eigenstates described by some Hamiltonian $H^{S}_{j,k',L,m}.$
         If the rational number states of the hybrid system,
         $S_{j,k',L,m},$ are energy eigenstates,
          then one has the eigenvalue equation\begin{equation}
          \label{SHenergy}H^{S}_{j,k',L,m}|\g,s\rangle_{j,k',L,m}
          =E(\g,s)_{j,k',L,m}|\g,s\rangle_{j,k',L,m}\end{equation}
          where $E(\g,s)_{j,k',L,m}$ is the energy eigenvalue of the
          state $|\g,s\rangle_{j,k',L,m}.$ The superscript $S$ on
          $H^{S}_{j,k',L,m}$  allows for the  possibility that the
          Hamiltonian depends on the type of hybrid system.

          The gauge variable has been removed because  the requirement
          that Eq. \ref{SHenergy} is satisfied for some choice of
          $H^{S}_{j,k',L,m},$ fixes the gauge or basis to be the
          eigenstates of $H^{S}_{j,k',L,m}.$  Since $H^{S}_{j,k',L,m}$
          is not known, neither is the dependence of $E(\g,s)_{j,k',L,m}$
          on $\g,s.$

          The existence of a Hamiltonian for the $S$ hybrid systems
          means that  there is energy associated with the values of
          rational numbers represented as states of hybrid systems.
          From this it follows that there are potentially
          many different energies associated with each rational
          number value.  This is a consequence of the fact that
          each rational number value has many string state
          representations that differ by the number of leading
          and trailing $0s.$

          One way to resolve this problem is to let the
          energy of a hybrid system state with no leading or
          trailing $0s$ be the energy value for the rational number
          represented by the state. In this way one has, for each
          $k,$ a unique energy associated with the value of the
          rational number shown by the state.

          One consequence of this association of energy to rational
          number values is that to each Cauchy sequence of rational
          number states of hybrid systems there corresponds a sequence
          of energies. The energy of the $nth$ state in the sequence
          is given by $\langle\g_{n},s_{n}|H_{j,k',L_{n},m_{n}}
          |\g_{n},s_{n}\rangle_{j,k',L_{n},m_{n}}.$

          It is not known at this point if the sequence of energies
          associated with a Cauchy sequence of hybrid system states
          converges or not. Even if energy sequences converge for
          Cauchy sequences in an equivalence class, the question
          remains whether or not the energy sequences converge to the
          same limit for all sequences in the equivalence class.

         The above description is valid for one hybrid system.  In
         order to describe more than one of these systems, another
         parameter, $h,$ is needed whose values distinguish the states of
         the different $S_{j,k',L,m}$ systems.  To this
         end the states $|\g,s\rangle_{j,k',L,m,g}$ of a system are
         expanded by including a parameter $h$ as in
         $|\g,h,s\rangle_{j,k',L,m,g}.$ In this case the state of two
         $S_{j,k',L,m}$ is given by $$|\g_{1},h_{1},s_{1},
         \rangle_{j,k',L,m,g'}|\g_{2},h_{2},s_{2}\rangle_{j,k',L,m,g'}$$ where
         $h_{1}\neq h_{2}$. This allows for the states of the two
         systems to have the same $\g$ and $s$ values.

         Pairs of hybrid systems are of special interest because
         states of these pairs correspond to  values of complex
         rational numbers. The state of one of the pairs is the real
         component and the other is the imaginary component. Since
         these components have different mathematical properties,
         the corresponding states in the pairs of states of hybrid
         systems must be distinguished in some way.

         One method is to distinguish the hybrid systems in the
         pairs by an index $r,i$ added to
         $S_{j,k,L,m}$ as in $(S_{r},S_{i})_{j,k',L,m}.$
         In this case states of $S_{r}$ and $S_{i}$ are values of
         the real and imaginary components of rational numbers. In
         this case complex numbers are Cauchy sequences of states of
         pairs, $(S_{r},S_{i})_{j,k',L_{n},m_{n}}$ of hybrid systems.

          As might be expected, the kinematics and dynamics of  hybrid
          systems $S_{j,k',L,m}$ in a frame $F_{j,k,g}$ are described
          relative to a space  and time lattice in the frame. For
          example a Schr\"{o}dinger equation description of two
          hybrid systems interacting with one another is given by
           \begin{equation}\label{Schreq} i\Delta^{f}_{t}
           \psi(t) =H\psi(t). \end{equation} $\Delta^{f}_{t}$
           is the discrete forward time derivative where
           $\Delta^{f}_{t}\psi(t)=(\psi(t+\Delta)-\psi(t))/\Delta.$
           Here $\psi(t)$ is the state of the two hybrid systems
           at time $t.$

           The Hamiltonian can be expressed as the sum of a
           Hamiltonian for the separate systems and an interaction
           part as in\begin{equation}\label{H0Hint}H=H_{0}+H_{int}.
           \end{equation} For two hybrid systems $H_{0}=\sum_{i=1}^{2}
           H_{0,i}$ where\begin{equation}\label{H0i}
           H_{0,i}
           =\frac{-\hbar^{2}}{2m_{S_{j,k',L,m}}}\Delta^{f}_{i}
           \cdot\Delta^{b}_{i}+H_{i,j,k',L,m}.\end{equation} The first
           term of $H_{0,i}$ is the  kinetic energy operator for the $ith$
           $S_{j,k',L,m}$ system.  The second term is the
           Hamiltonian for the internal states of the system. It is given
           by Eq. \ref{SHenergy}. Also $m_{S_{j,k',L,m}}$ is the mass
           of $S_{j,k',L,m}$. $\Delta^{f}$ and $\Delta^{b}$ are the
           discrete forward and backward space derivatives, and
           $\hbar$ is Planck's constant. The dot product indicates the
           usual sum over the product of the $D$ components in the
           derivatives. Note that a possible dependence of the mass
           of $S_{j,k',L,m}$ on $k',L,m$ has been included.

           The question arises regarding how one should view
           $N$-tuple hybrid systems as physical systems.  Should they
           be regarded as $N$ independent systems each with its own
           Hamiltonian ($H_{int}=0$ in Eq. \ref{H0Hint}) or as systems
           bound together with energy eigenstates that are quite
           different from those of the single hybrid systems in isolation.

           One way to shed light on this question is to examine
           physical representations of number tuples
           in computers. There $N$-tuples of numbers are represented as
           $N$ strings of bits or of qubits (spin $1/2$ systems)
           bound to a background matrix of potential wells where
           each well contains one qubit. The locations of the qubits
           in the background matrix determines their assembly into
           strings and into tuples of strings.

           Here it is assumed that $N$ tuples of hybrid systems
           consist of $N$ $S_{j,k',L,m}$ systems bound
           together in some fashion.  Details of the binding, and its
           effect on the states of the individual $S_{j,k',L,m}$
           in the $N$-tuple are not known at this point. However it
           will be assumed that the effect is negligible. In this
           case the energy of each component state
           $|\g_{z},h_{z},s_{z}\rangle$ in the $N$ tuple will be
           assumed to be the same as that for an individual $S_{j,k',L,m}$
           system. Then, the energy of the state $|\bar{\g},\bar{h},
           \bar{s}\rangle$ is the sum of the energies of the individual
           component states. Also the energy is assumed to be
           independent of the $h$ values.

           This picture is supported by the actual states of computers
           and their computations.  The background potential well
           matrix that contains the $N$-tuples of qubit string
           states is tied to the computer.  Since the computer
           itself is a physical system, it can be translated, rotated,
           or given a constant velocity boost.
           In all these transformations the states of the qubit strings
           in the $N$-tuples and the space relations of the $N$ qubit
           strings to one another is unchanged.  These parameters
           would be changed if two computers collided with one
           another with sufficient energy to disrupt the internal
           workings.

           This picture of each frame containing physical systems
           and a plethora of different hybrid systems and their tuples
           may seem objectionable.  However, one should recall that
           here one is working in a possible domain of a coherent
           theory of physics and mathematics together. In this case
           the domain might be expected to include many types of
           hybrid systems that have both physical and mathematical
           properties. This is in addition to the presence of
           physical systems and  mathematical systems.

           \section{Frame Entities as Viewed in a Parent Frame}\label{FEVPF}

          So far it has been seen that each frame, $F_{j,k,g},$ in
          the frame field contains a set $\mathcal{L}_{j,k}$ of space
          and time lattices where the number of points in each dimension
          and the point spacing satisfy Eq. \ref{Mdelta} for some $L$ and
          $m$. The frame also contains qukit string systems as hybrid
          $S_{j,k',L',m'}$ systems and various tuples of these
          systems.  Here $k',L',m'$ need have no relation to $k,L,m.$ Each
          frame also contains  physical theories  as mathematical
          structures based on $R_{j,k,g},C_{j,k,g}$ the real and
          complex number base of the frame. These theories describe
          the kinematics and dynamics of physical systems on the
          space and time lattices in the frame. For quantum systems
          these theories include Hilbert spaces as vector spaces
          over $C_{j,k,g}.$

          Since this is true for every frame, it is true for a
          frame $F_{j,k,g}$ and for a parent frame $F_{j-1,k',g'}.$
          This raises the question of how entities in a frame are
          seen by an observer in a frame at an adjacent iteration
          stage.  As was noted in Section \ref{FIRF}, it is assumed
          that ancestor frames and their contents are not visible to
          observers in descendant frames; however, descendant frames
          and their contents are visible to observers in ancestor
          frames.  It follows that observers in frame $F_{j,k,g}$
          cannot see frame $F_{j-1,k',g'}$ or any of its contents.
          However, observers in $F_{j-1,k',g'}$ can see $F_{j,k,g}$
          and its contents.\footnote{This restriction has to be
          relaxed for cyclic frame fields.}

          One consequence of the relations between frames at
          different iteration stages is that entities
          in a frame, that are seen by an observer in the
          frame as featureless and with no structure, correspond
          to entities in a parent frame that have structure.
          For example, elements of $R_{j,k,g},C_{j,k,g}$ are seen
          by an observer in $F_{j,k,g}$ as abstract, featureless objects
          with no properties other than those derived from the relevant
          axiom sets. However, to observers in a parent frame
          $F_{j-1,k',g'},$ numbers in $R_{j,k,g},C_{j,k,g}$  are seen
          as equivalence classes (pairs of equivalence classes for
          $C_{j,k,g}$)  of Cauchy sequences of states of base $k$
          qukit string systems.  Thus entities that are abstract and featureless
          in a frame have structure as elements of a parent frame.

           It is useful to represent these two in-frame views by
         superscripts $j$ and $j-1$. Thus $R^{j-1}_{j,k,g},
         C^{j-1}_{j,k,g}$ and $R^{j}_{j,k,g},C^{j}_{j,k,g}=R_{j,k,g},C_{j,k,g}$
         denote the stage $j-1$ and stage $j$ frame views of the number
         base of frame $F_{j,k,g}.$ They are often referred to in the
         following as parent frame images of $R_{j,k,g},C_{j,k,g}$.

          The distinction between elements of a frame and their images in
          a parent frame exists for other frame entities as well.
          The state\begin{equation}\label{psija}\psi_{j}=\sum_{\a}
          d^{j}_{j,\a}|\a\rangle^{j}_{j,k,g}\end{equation} in
          $\mathcal{H}^{j}_{j,k,g}$ corresponds to the state
          \begin{equation}\label{psijj-1a}\psi^{j-1}_{j}=
          \sum_{\a}d^{j-1}_{j,\a}|\a\rangle^{j-1}_{j, k,g}\end{equation}
          in $\mathcal{H}^{j-1}_{j,k,g},$ which is the parent frame
          image of $\mathcal{H}_{j,k,g}.$  In the above $d^{j}_{j,\a}$
          is a featureless abstract complex number in $C_{j,k,g}$ whereas
          $d^{j-1}_{j,\a},$ as an element of $C^{j-1}_{j,k,g},$ is an
          equivalence class of Cauchy sequences of hybrid system states.

          The use of stage superscripts and subscripts applies to
          other frame entities, such as hybrid systems, physical
          systems, and space and time lattices.A hybrid system
          $S_{j,k',L',m'}$ in $F_{j,k,g}$ has $S^{j-1}_{j,k',L',m'}$ as
          a parent frame image.   States $|\g',s'\rangle_{j}$ of $S_{j,k',L',m'}$ are
          vectors in the Hilbert space $\mathcal{H}_{j,k,g},$  in
          $F_{j,k,g}.$ States $|\g',s'\rangle^{j-1}_{j}$
          of $S^{j-1}_{j,k',L',m'}$ are vectors in
          $\mathcal{H}^{j-1}_{j,k,g}.$  The two
          states are different in that $|\g',s'\rangle_{j}$ is
          an eigenstate of an operator whose corresponding eigenvalue
          $\g',s'$ is a rational real number with no structure.

          The state $|\g',s'\rangle^{j-1}_{j}$ of $S^{j-1}_{j,k',L',m'}$ is
          an eigenstate of a number value  operator whose corresponding eigenvalues
          are rational real numbers in $R^{j-1}_{j,k,g}$ corresponding to
          $(\g',s').$ These eigenvalues are equivalence classes of
          Cauchy sequences of states of hybrid systems  $S_{j-1,k,L,m}$ in
          a $j-1$ stage frame. Here $k,j-1$ are fixed and $L=L_{n}$ and
          $m=m_{n}$  for the $nth$ term in the sequence.
          This accounts for the fact that  each state in the sequence
          is a state of a different hybrid system with $j-1,k$ held fixed.

         The eigenvalue equivalence class is a base $k'$ real rational number.
         Since it is an element of $R^{j-1}_{j,k,g}$,  it contains a constant
         sequence of hybrid system states if and only if all prime factors of
         $k'$ are factors of $k.$ If this is the case then one can equate
         the equivalence class to a single state of a hybrid system to conclude
         that the eigenvalue associated with $|\g',s'\rangle^{j-1}_{j}$
          is a state of a hybrid system $S_{j-1,k,L,m}$ in stage $j-1$
          frame.\footnote{Note that the
          language used here avoids referring to the absolute
          existence of  systems with properties independent of the
          frames.  The emphasis is on the view of systems in different
          frames. Thus $b_{j}=b^{j}_{j}$ is the view of a physical
          system $b$ as seen by an observer in a stage $j$ frame.
          The image of this view in a stage $j-1$ frame is denoted by
          $b^{j-1}_{j}.$ The difference between the two is that
          physical properties of $b_{j},$ as eigenvalues of operators
          over $\mathcal{H}_{j,k,g},$ are featureless abstract real
          numbers. Properties of $b^{j-1}_{j},$ as operator eigenvalues,
          are equivalence classes of Cauchy sequences of hybrid system states.
          A frame independent description would be expected to appear only
          asymptotically when the frames in the field are merged.}

          If $k'$ has prime factors that are not prime factors of $k$
          (such as $k'=6$ and $k=3$),  the eigenvalue for the eigenstate
          $|\g',s'\rangle^{j-1}_{j}$ is still a real rational number.
          However, as an equivalence class of Cauchy sequences
          of base $k$ hybrid system states,\footnote{which it
          must be as a real number in $C^{j-1}_{j,k,g}$}  it does not
          contain a constant sequence of hybrid system states. Instead it
          contains a sequence that corresponds to an infinite repetition of
          a base $k$ hybrid system state (Just like the decimal expansion of
          $1/6=0.16666\cdots$).

          \subsection{Parent Frame Views of Lattice Point Locations}
          \label{PFVLPL}

          A similar representation holds for parent frame views
          of  point locations of lattices.  Let
          $L_{j,k,L,m}$ denote a lattice of $D$ space dimensions
          and one time dimension in a frame $F_{j,k,g}.$ The
          components $x_{j,z}$ (with $z=1,\cdots ,D$) of the $D$ space
          locations $\bar{x}_{j}$ of the lattice points, $p_{j},$ are such that
          $x_{j,z}$  is a rational real number. The lattice
          points are abstract and have no structure other than that
          imparted by the values $\bar{x}_{j}.$ As rational real numbers
          in $R_{j,k,g},$ the $x_{j,z}$ have no structure other than the
          requirement that they are both rational and real numbers.

          The view or image of $L_{j,k,L,m}$ from
          the position of an observer in a stage $j-1$ parent frame
          is denoted by $L^{j-1}_{j,k,L,m}.$  The image
          points and locations of points in $L^{j-1}_{j,k,L,m}$ are
          denoted by $p^{j-1}_{j}$ and $\bar{x}^{j-1}_{j}.$

          The space point locations $\bar{x}^{j-1}_{j}$ are different
          from the $\bar{x}_{j}$ in that they have more structure. This
          follows from the fact that they are $D$ tuples of rational
          real numbers in $R^{j-1}_{j,k,g}.$  It follows from this
          that each component, $x^{j-1}_{j,z},$  of a space
          point location, $\bar{x}^{j-1}_{j},$ in $L^{j-1}_{j,k,L,m}$ is
          an equivalence class of Cauchy sequences
          of states of hybrid systems. Since each of the $D$ equivalence
          classes is a real number equivalent
          of a rational number value, the equivalence class
          includes many (numerically) constant sequences
          of hybrid system states.  The existence of constant sequences
          follows from the observation that the $k$ subscript of the
          lattice image is the same as that for $R^{j-1}_{j,k,g}$.
          The states in the different sequences differ by the presence
          of different numbers of leading and trailing $0s.$

          A useful way to select a unique constant sequence is to
          require that all states in the sequence be a unique state
          of the hybrid system  $S_{j-1,k,L,m}$ where the $k,L,m$
          subscripts are the same as those for $L^{j-1}_{j,k,L,m}.$
          Replacement of the sequence by its single component state
          gives the result that, for each $z,$ $x^{j-1}_{j,z}$ is a
          state of  $S_{j-1,k,L,m}.$

          It follows from this that the locations of space points
          $p_{j}$ in $L_{j,k,L,m},$ as viewed from a stage $j-1$ frame, are
          seen as states of a $D$-tuple $S^{D}_{j-1,k,L,m}$ of hybrid
          systems in the stage $j-1$ frame. These states correspond to
          $D$-tuples of rational number values.

          A similar representation of the time points in the
          lattices is possible for the  real rational
          number time values.  In this case the time values of a lattice
          are seen in a parent frame as rational number states of a
          hybrid system.

          \section{Lattices and Hybrid Systems}\label{LHS}
          The above description of a stage $j-1$ frame view of lattices
          gives point locations of $L^{j-1}_{j,k,L,m}$ as states of a $D$ tuple of
          hybrid systems $S^{x,D}_{j-1,k,L,m}$ for the space part and
          states of another hybrid system $S^{t}_{j-1,k,L,m}$ for the time
          part. The superscripts $s$ and $t$ denote the hybrid
          systems associated with space and time points
          respectively.

          This strongly suggests that each space point image
          $p^{x,j-1}_{j}$ in the space part of a parent
          frame image, $L^{j-1}_{j,k,L,m},$ of $L_{j,k,L,m}$ be
          identified with a $D$ tuple, $S^{x,D}_{j-1,k,L,m}$ of
          hybrid systems and each time point image $p^{t,j-1}_{j}$ in the time
          part be identified with a single hybrid system, $S^{t}_{j-1,k,L,m}.$
          For each image point $p^{x,j-1}_{j},$ the location is given by the
          state of the  $D$ tuple $S^{x,D}_{j-1,k,L,m},$  of  hybrid systems
          in a parent frame that is associated with the space point image.
          Similarly the location of each image time point in $L^{j-1}_{j,k,L,m}$
          is given by the state of the hybrid system, $S^{t}_{j-1,k,L,m},$
          associated with the point.

          This shows that set of all parent frame images of the $k^{LD}$ space
          points of $L_{j,k,L,m}$ become  a set of $k^{LD}$ $D$ tuples of parent frame
          hybrid systems, $S^{x,D}_{j-1,k,L,m},$ with the state of each $D$ tuple
          corresponding to the image space point location in $L^{j-1}_{j,k,L,m}.$
          The parent frame images of the $k^{D}$ time points of $L_{j,k,L,m}$
          become  a set of  $k^{L}$ hybrid systems $S^{t}_{j-1,k,L,m}.$ Each is
          in a different state corresponding to the different possible lattice
          time values.

\begin{figure}[h]
         \begin{center} \resizebox{150pt}{150pt}{\includegraphics
         [240pt,120pt][580pt,560pt]{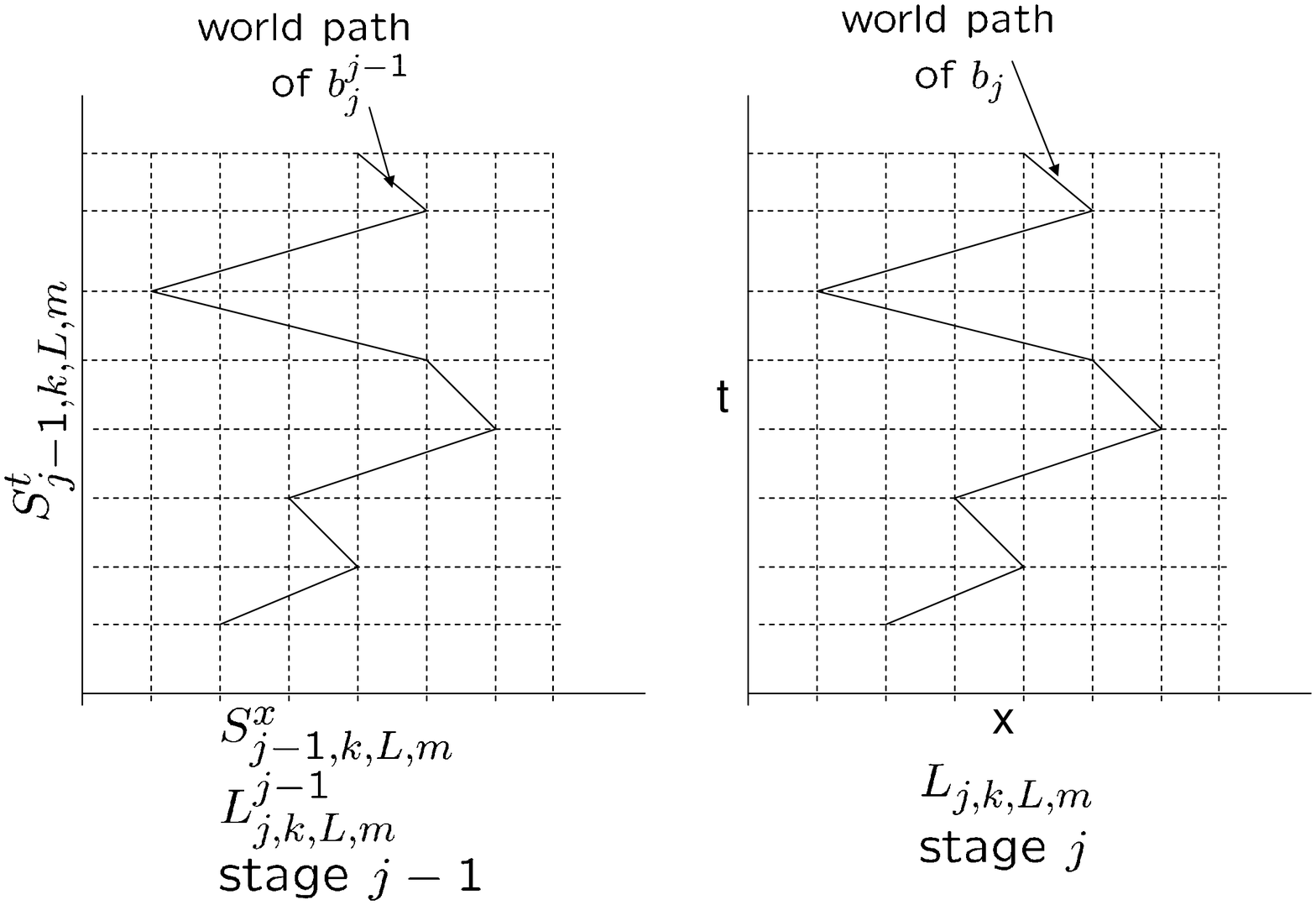}}\end{center}
         \caption{Stage $j$ and stage $j-1$ images of one dimensional
         space and time lattices. Lattice points are indicated by
         intersections of lines in the two dimensional  grid.
         In $L_{j,k,L,m}$ the points $p_{j}$ consist of
         pairs of rational numbers. In $L^{j-1}_{j,k,L,m}$ the image
         points $p^{j-1}_{j}$ consist of pairs $S^{x}_{j-1,k,L,m},
         S^{t}_{j-1,k,L,m}$ of hybrid systems with with point
         locations given by the states of the system pairs. Non
         relativistic world paths of a stage $j$ physical system
         $b_{j}$ and of its stage $j-1$ image  $b^{j-1}_{j}$
         are shown as solid lines. Note that $b$ can also be
         a hybrid system.}\label{PACTPM3} \end{figure}

          Figure \ref{PACTPM3} illustrates the situation described above for
          lattices with one space and one time dimension. The stage
          $j$ and $j-1 $ lattice points are shown by the
          intersection of the grid lines. For the stage $j$ lattice
          the points correspond to rational number pairs whose
          locations are given by the values of the pairs of numbers.
          For the stage $j-1$ image lattice the points correspond to
          pairs of hybrid systems, one for the space dimension and
          one for the time dimension. Point locations are given by
          the states of the hybrid system pairs.  Non relativistic
          world paths for physical systems $b_{j}$ and its stage
          $j-1$ image $b^{j-1}_{j}$ are also shown.

           It is of interest to compare the view here with that in
           \cite{Tegmark}. Tegmark's explicitly stated view is that
           real numbers, as labels of space and time points, are distinct
           from the points themselves. This is similar to the setup here in
           that points of parent frame images of lattices are  tuples of hybrid
           systems, and locations or labels of the points of parent frame
           images of lattice are states of tuples of hybrid systems.

           The differing  views of hybrid systems as either
           number systems or physical systems may seem strange
           when viewed from a perspective outside the frame
           field and in the usual physical universe. However it is
           appropriate for a coherent theory of physics and mathematics
           together as such a theory might have systems that represent
           different entities, depending  on how they are viewed.

          \subsection{Energy of Space Points in $L^{j-1}_{j,k,L,m}$}

          The description of parent frame images of lattice space points
          and their locations  as $D$ tuples of hybrid systems and
          states of the tuples, means that the image of each point has
          a mass. The mass is equal to that of the $D$ tuple of hybrid
          systems associated with each point image.  The (rest) masses
          of all space points in an image lattice $L^{j-1}_{j,k,L,m}$
          should be the same as the $D$ tuples of hybrid systems associated
          with each point are the all the same. However each of
          the tuples is in a different image state $\bar{x}^{j-1}_{j}$ that
          corresponds to the different locations of each point
          image.

           Each component $x^{j-1}_{j,z}$ of $\bar{x}^{j-1}_{j}$ corresponds
           to a hybrid system state $|+,s_{z}\rangle^{j-1}_{j}$ of a
           component hybrid system $S^{D}_{z,j-1,k,L,m}$ in the $D$ tuple
           (note the subscript $z$). Each of these component hybrid
           system states is an energy eigenstate of a hybrid
          system Hamiltonian.  The corresponding energy eigenvalue,
          $E(+,s_{z})_{j-1},$ is defined by Eq. \ref{SHenergy}.
          Here $\g$ is positive as the lattice
          component locations are all $\geq 0.$

          If the component hybrid systems in a $D$ tuple do not
          interact with one another, then the energy associated with
          a parent frame image lattice location is the sum of the
          component energies. In this case the energy associated
          with the location image, $|+,\bar{s}\rangle^{j-1}_{j}$, of
          $\bar{x}_{j}$ is given by\begin{equation}\label{energys}
          E(+,\bar{s})_{j-1}=\sum_{z=1}^{D}E(+,s_{z})_{j-1}.
          \end{equation}Here $|+,\bar{s}\rangle^{j-1}_{j}$ denotes
          the tensor product of all $D$ component states
          $|+,s_{z}\rangle^{j-1}_{j}.$ If the component systems in a
          $D$ tuple do interact with one another then the situation
          becomes more complex as the Hamiltonian and energy eigenvalues
          must take account of the interactions.

          At this point the specific dependence of the energy on the
          parent frame image of the lattice point locations
          is not known as it depends on the properties of the hybrid
          system Hamiltonian. Nevertheless the  existence of energies
          associated with  locations of points of parent frame images of a  space
          lattice is intriguing. One should note, though, that this
          association of energy to space points holds only for
          parent frame images.  It does not extend to lattices
          $L_{j,k,L,m}$ when viewed by an observer in $F_{j,k,g}.$

          This aspect is one reason why one needs to do more work,
          particularly on the merging of frames in the frame field.
          It is quite possible that, in the case of a cyclic frame field,
          some aspects of the association of energy with space
          points in the same frame will be preserved. Note that for
          cyclic frame fields the restriction that an observer
          cannot see ancestor frames or their contents must be
          relaxed.  The reason is that ancestor frames are also
          descendant frames.

           \subsection{Dynamics of  Systems as Seen from
           a Parent Frame}\label{DSSPF}

           The description of the motion of hybrid systems and other
           physical systems in a stage $j$ frame, as seen from a
           parent stage $j-1$ frame, is interesting. The reason is
           that the dynamics and kinematics of systems are based on a
           parent frame image  lattice, $L^{j-1}_{j,k,L,m},$ whose
           space points  and point locations are $D$ tuples,
           $S^{x,D}_{j-1,k,L,m}$, and tensor product states
           $\bigotimes_{z=1}^{D}|+,s_{z}\rangle_{j-1}$ of the
           $D$ tuples of hybrid systems. The time points and point
           locations are hybrid systems, $S^{t}_{j-1,k,L,m}$ and
           states of these systems. The $x$ and $t$ superscripts
           allow for the possibility that the hybrid systems associated
           with space point images may be different from those associated
           with time images.

           It follows that the stage $j-1$ frame description of the motion
           and dynamics of stage $j$ systems is described relative to the
           states of certain hybrid systems in the parent frame.
           To understand this consider a simple  example where
           $\Psi(x,t)_{j}$ denotes the state of some physical
           system $b_{j}$ in a stage $j$ frame at position $x$ and at time $t.$
           The pair $x,t$ denote a point in a  stage $j$ frame space and
           time lattice $L_{j,k,L,m}$ which, for simplicity, consists of
           one space and one time dimension.  The stage $j$ frame time
           evolution of the state  $\Psi(x,t)_{j}$ is given by a discrete
           Schr\"{o}dinger Equation,\begin{equation}\label{Schrj}i\hbar
           \Delta^{f}_{t,j}\Psi_{j}(x,t)=H_{j}\Psi_{j}(x,t).\end{equation}
            Here $\Delta^{f}_{t,j}$ is the discrete forward time
            derivative defined by\begin{equation}\label{Deltatn}
            \Delta^{f}_{t,j}\Psi_{j}(x,t)=\frac{\Psi_{j}
            (x,t+\Delta_{j})-\Psi_{j}(x,t)}{\Delta_{j}}.\end{equation}

            As a sum of kinetic and potential terms, the Hamiltonian,
            $H_{j},$ has the form\begin{equation}\label{Hamj}H_{j}=
            -\frac{\hbar^{2}\Delta^{f}_{j}\Delta^{b}_{j}}{2m_{b_{j}}}+V_{j}.
            \end{equation} To keep things simple, the
            description is restricted to just one $b_{j}$ system
            interacting with an external potential. In this case
            $V_{j}=V_{j}(x)$  Also $\hbar$ is Planck's constant
            and $m_{b_{j}}$ is the mass of system $b_{j}$.

            In the above, the forward and backward discrete derivatives
            $\Delta^{f}_{j}$ and $\Delta^{b}_{j}$ are defined
            similar to the forward time derivative. One has
            \begin{equation}\label{Deltafb}\begin{array}{c}
            \Delta^{f}_{j}\Psi_{j}(x,t)=\frac{\Psi_{j}(x+\Delta_{j},t)-
            \Psi_{j}(x,t)}{\Delta_{j}}\\\\\Delta^{b}_{j}\Psi_{j}(x,t)=
            \frac{\Psi_{j}(x-\Delta_{j},t)-\Psi_{j}(x,t)}{\Delta_{j}}.
            \end{array}\end{equation}

            This description of the time development and Hamiltonian for
            a system $b_{j}$ is a description in a stage $j$ frame.
            Viewed from a parent stage $j-1$ frame  the image of the
            Schr\"{o}dinger equation, Eq. \ref{Schrj} describes the motion
            of the system $b$ in the image lattice $L^{j-1}_{j,k,L,m}.$
            Since the space and time point locations in the image
            lattice are states of hybrid systems $S^{x}_{j-1,k,L,m}$ and
            $S^{t}_{j-1,k,L,m}$, the image Schr\"{o}dinger equation
            describes the motion  of system $b$ relative to these
            states. The image equation is given by\begin{equation}
            \label{Schrjj-1}i\hbar(\Delta^{f})^{j-1}_{j,s_{t}}\Psi^{j-1}_{j}
            (s_{x},s_{t})=H^{j-1}_{j}\Psi^{j-1}_{j}(s_{x},s_{t}).
            \end{equation} The image state, $\Psi^{j-1}_{j}(s_{x},s_{t}),$
            is the same state in the Hilbert space
            $\mathcal{H}^{j-1}_{j,k,g}$ as $\Psi_{j}(x,t)$ is in
            $\mathcal{H}_{j,k,g}.$ Here $s_{x}$ and $s_{t}$ are
            shorthand notations for the hybrid system states,
            $|+,s_{x}\rangle_{j-1}$ and $|+,s_{t}\rangle_{j-1}$ of
            $S^{x}_{j-1,k,L,m}$ and $S^{t}_{j-1,k,L,m}$

            The Hamiltonian $H^{j-1}_{j}$ is given by \begin{equation}
            \label{Hamjj-1}H^{j-1}_{j}=-\frac{\hbar^{2}(\Delta^{f})^{j-1}_{j,s_{x}}
            (\Delta^{b})^{j-1}_{j,s_{x}}}{2m_{b^{j-1}_{j}}}+V^{j-1}_{j}(s_{x}).
            \end{equation} The potential $V^{j-1}_{j}(s_{x})$ is a function
            of the states $|+,s_{x}\rangle_{j-1}$ of $S^{x}_{j-1,k,L,m}.$ The value of
            $m_{b^{j-1}_{j}}$ is a real number in $R^{j-1}_{j,k,g}$
            that is expected to be the same as that of $m_{b^{j}_{j}}$
            in $R^{j}_{j,k,g}.$

            The  forward and backward discrete
            derivatives are expressed by equations similar to Eq.
            \ref{Deltafb}:\begin{equation}\label{Deltafbj-1}
            \begin{array}{c}(\Delta^{f})^{j-1}_{j,s_{x}}
            \Psi^{j-1}_{j}(s_{x},s_{t})=\frac{\Psi^{j-1}_{j}(s_{x}+1,s_{t})
            -\Psi^{j-1}_{j}(s_{x},s_{t})}{\Delta^{j-1}_{j}}
            \\\\(\Delta^{b})^{j-1}_{j,s_{x}}\Psi^{j-1}_{j}(s_{x},s_{t})
            =\frac{\Psi^{j-1}_{j}
            (s_{x}-1,s_{t})-\Psi^{j-1}_{j}(s_{x},s_{t})}{\Delta^{j-1}_{j}}.
            \end{array}\end{equation} In these equations $s_{x}+1$
            and $s_{x}-1$ denote the hybrid system states
            $|+,s_{x}+_{A}1\rangle_{x}$ and $|+,s_{x}-_{A}1\rangle_{x}$
            where the subscript $A$ denotes arithmetic addition and
            subtraction. For example if $|+,s_{x}\rangle_{x}=
            |100+111\rangle_{x}$ in binary then $|+,s_{x}+_{A}1
            \rangle_{x}=|101+000\rangle_{x}.$

            This use of states of $S^{x}_{j-1,k,L,m}$ and
            $S^{t}_{j-1,k,L,m}$ as point locations of an image space
            and time lattice must be reconciled with the observation
            that these hybrid systems are dynamical systems that move
            and interact with one another and with other physical
            systems. If the $S^{x}_{j-1,k,L,m}$ and
            $S^{t}_{j-1,k,L,m}$ and their states are to serve as
            points and point locations of a space and time lattice
            image, they must be dynamically very stable and
            resistant to change.  This suggests that these systems
            must be very massive and that their interactions with each
            other and other systems is such that state changes occur
            very rarely, possibly on the order of cosmological time
            intervals.\footnote{The  dynamics of these systems on
            space and time lattices $L_{j-1,k',L',m'}$ is described
            in a stage $j-1$ frame. Here $k',L',m'$ need have no
            relation to $k,L,m.$}

            The reason for these restrictions on the properties of
            space and time hybrid systems is that one expects the
            space and time used to describe motion of systems to
            be quite stable and to change at most very slowly.
            Changes, if any, would be expected to be similar to
            those predicted by the Einstein equations of general
            relativity.

            \section{Discussion}\label{D}

            It is to be emphasized that the work presented so far is
            only a beginning to the development of a complete framework
            for a coherent theory of physics and mathematics together.
            Not only that but one must also find a way to reconcile
            the multiplicity of universes, one for each frame, to
            the view that there is only one physical universe.

            One way to achieve reconciliation is to drop the single
            universe view and to relate the
            multiplicity of frame representations of physics and
            mathematics to the different many physical universes view
            of physics. These include physical universes in existing
            in different bubbles of space time \cite{Linde,Susskind},
            and other descriptions of multiple universes
            \cite{Davies,Davies1,Tegmark2,Deutsch} including the
            Everett Wheeler description \cite{EW}. Whether any of these
            are relevant here or not will have to await future work.

            If one sticks to the single physical universe view then
            the frames with their different universes and space and
            time representations need to be merged or collapsed to
            arrive in some limit at the
            existing physical universe with one space time. This
            applies in particular to the iteration stage and gauge
            degrees of freedom as their presence is limited to the
            quantum representation of numbers.

            One expected consequence of the merging is that it
            will result in the emergence of a single background
           space time as an asymptotic limit of the merging of the space
           time lattices in the different frames. Whether the ultimate space
           time background is a continuum, a foam \cite{Ng,Ng1,Amelino,Kirillov},
           or has some other form, should be determined by details of
           the merging process.

           In addition the merging may affect other entities
           in the frames.  Physical systems, denoted collectively
           as $b_{j}$, may become the  observed physical systems
           in the space time background. In addition, hybrid systems
           may split into either physical systems or mathematical
           systems.

           One potentially useful approach to frame merging is the
           use of  gauge theory techniques \cite{Montavy,Enk} to
           merge frames in different iteration stages.  One hopes
           that some aspects of the standard model \cite{STDMOD}
           in physics will be useful here. This will require
           inclusion of relativistic treatment of systems  and
           quantum fields in the frames.

            This look into a possible future approach emphasizes
            how much there is to accomplish. Nevertheless one may
            hope that the work presented here
            is a beginning to the development of a  coherent theory.
            The expansion of the frames in the frame field to
            include, not only mathematical systems, but also space
            and time lattices, and hybrid systems that are both mathematical
            systems and physical systems, seems reasonable from the
            viewpoint of a coherent theory of physics and
            mathematics together.  One might expect such a theory to
            contain systems that can be either physical systems or
            mathematical systems.

            The use of massive hybrid systems to be stage $j-1$ frame images
            of points of space and time lattices in stage $j$ frames
            suggests that there must be different types of hybrid systems.
            For example, stage $j$ theoretical predictions of the values
            of some physical quantity $Q$ are, in general, real numbers in
            $R_{j,k,g}.$ Their images in a stage $j-1$ parent frame are
            equivalence classes of Cauchy sequences of states of hybrid systems
            $S_{j-1,k,L,m}$ where $L,m$ are dependent on positions in the Cauchy
            sequences.\footnote{If one introduces limit hybrid systems
            containing an infinite number of qubits, then an equivalence
            class of Cauchy sequences could be replaced by just one limit
            system whose state corresponds to an infinite sequence of digits
            that is a canonical representative of the class.} If the predicted
            values are rational numbers expressible as a finite string of
            base $k$ digits, then the stage $j-1$ values can be
            expressed as states of $S_{j-1,k,L,m}$ rather than equivalence
            classes of sequences of states. If the images of properties of the
            physical quantities $Q$ are to be reflected in the properties of
            the hybrid systems, then different types of hybrid systems, such as
            $S^{Q}_{j-1,k,L,m},$ must be associated with different physical
            quantities.

            Whether these descriptions of parent frame images as
            hybrid systems will remain or will have to be modified,
            remains to be seen. However, it should be recalled that
            these images are based on the dual role played by values
            of rational numbers both as mathematical systems and as
            locations of components of points in the lattices.
            Recall that the notion of a point in a lattice is
            separated from the location of the point just as the
            notion of a number, as a mathematical system, is separated
            from the value of a number. This use of number and
            number value is different from the usual use in
            mathematics in that expressions, as $135.79,$ are
            usually considered as rational numbers and not as
            values of rational numbers. Here a rational number,
            as a hybrid system, is similar to the usual mathematical
            concept of a set of rational numbers as a model of the
            rational number axioms.

           In conclusion, it is worth reiterating the last paragraphs
           at the end of the introduction. Whatever one thinks of
           the ideas presented in the paper, the following points should be
           kept in mind. Two of the three dimensions of the field of reference
           frames are present only for quantum theory representations
           of the real and complex numbers.  These are the gauge or
           basis degree of freedom and the iteration stage degree of
           freedom. They are not present in classical descriptions.
           The number base degree of freedom is present for both
           quantum and classical representations based on rational
           number representations by digit strings.

           The presence of the gauge and iteration degrees of
           freedom in the quantum representation described here is
           independent of the description of rational number values
           as states of qukit string systems. Any quantum
           representation of the rational numbers, such as states
           $|l,n\rangle$ of integer pairs, where the
           states are elements of a vector space will result
           in a frame field with gauge and iteration degrees of freedom.

           Finally, the importance of numbers to physics and
           mathematics should be emphasized. It is hoped that more
           work on combining quantum physics and the quantum theory
           of numbers will be done. The need for this is based on
           the observation that natural numbers, integers,
           rational numbers, and probably real and complex numbers,
           are even more fundamental to physics than is geometry.

               \section*{Acknowledgement}
            This work was supported by the U.S. Department of Energy,
          Office of Nuclear Physics, under Contract No.
          DE-AC02-06CH11357.

            \end{document}